
\magnification=1200
\baselineskip=20pt
\overfullrule=0pt
\tolerance=100000
\rightline{UR-1339}
\rightline{ER40685-788}
\smallskip

\centerline{\bf A NEW CLASS OF SUPERSYMMETRIC THEORIES}

\vskip 1.5cm

\centerline{Ashok Das}
\centerline{Department of Physics and Astronomy}
\centerline{University of Rochester}
\centerline{Rochester, NY 14627}

\vskip 3.5cm

\centerline{\bf \underbar{Abstract}}

We construct a class of quantum mechanical theories which are invariant
under fermionic transformations
 similar to supersymmetry transformations.
  The generators of the transformations in
this case, however, satisfy a BRST-like algebra.

\vskip 2in

\noindent (To be published in the Memorial Volume for R. E. Marshak.)

\vfill\eject

My acquaintance with Prof. R. E. Marshak was brief.  But during the few
meetings and discussions we had, he left a very strong impression on me.
He was bigger than life.  Study of symmetries
seemed to dominate his seemingly endless energy and interest in physics.
He always wanted to explore new symmetries and it is with this thought that
I dedicate this paper to Bob Marshak's memory.

It is well known that the quantum mechanical Lagrangian [1]
$$L = {1 \over  2}\ (\dot q)^2 - {1 \over 2}\ (f (q))^2 + i \overline \psi
\dot \psi - f^\prime (q) \overline \psi \psi \eqno(1)$$
where $f(q)$ is any monomial of the coordinates $q(t)$ and where $\psi$ and
$\overline \psi$ are fermionic variables is invariant under the
supersymmetry transformations
$$\eqalign{\delta q &= {1 \over \sqrt{2}} \ \overline \psi \epsilon\cr
\noalign{\vskip 4pt}%
\delta \psi &= - {i \over \sqrt{2}}\ \dot q \epsilon -
{1 \over \sqrt{2}}\ f(q) \epsilon \cr
\noalign{\vskip 4pt}%
\delta \overline \psi &= 0\cr}\eqno(2)$$
and
$$\eqalign{\overline \delta q &= {1 \over \sqrt{2}}\ \overline \epsilon
\psi\cr
\noalign{\vskip 4pt}%
\overline \delta \psi &= 0 \cr
\noalign{\vskip 4pt}%
\overline \delta \ \overline \psi &= {i \over \sqrt{2}}\ \dot q
\overline \epsilon - {1 \over \sqrt{2}}\ f (q) \overline \epsilon \cr}
\eqno(3)$$
Here $\epsilon$ and $\overline \epsilon$ are the constant Grassmann
parameters of the two supersymmetry transformations respectively.  It is
straightforward to check that
$$\left[ \delta_1 , \overline \delta_2 \right]
 \phi = - i \dot \phi \overline
\epsilon_2 \epsilon_1 \eqno(4)$$
for any variable $\phi$.  This is equivalent to saying that the two
supersymmetry charges satisfy the algebra [2].
$$\left[ Q , \overline Q \ \right]_+ = H \eqno(5)$$
where $H$ is the Hamiltonian of the system.  It is also known that if
$f(q)$ is an even monomial of $q$, then supersymmetry is spontaneously
broken through instanton effects.

\medskip

This is the usual description of supersymmetric quantum mechanics.  In what
follows, we will propose a new class of supersymmetric quantum mechanical
theories where the algebra is analogous to that of BRST algebra [3].  Let us
note that a nonrelativistic quantum mechanical system is naturally
described in terms of a Hamiltonian.  Let us consider the simple
Hamiltonian
$$H = {1 \over 2}\ p^2 + g p \overline \psi \psi \eqno(6)$$
where $p$ is the bosonic momentum variable and $g$ a coupling constant.
This is simply the Hamiltonian for a free bosonic particle interacting with
fermions through a \lq\lq derivative" coupling.  It is easy to check that
this Hamiltonian is invariant under the \lq\lq supersymmetry" transformations
$$\eqalign{\delta \psi &= - {i \over \sqrt{2}}\ \epsilon \cr
\noalign{\vskip 4pt}%
\delta \overline \psi &= 0 \cr
\noalign{\vskip 4pt}%
\delta p &= {i g \over \sqrt{2}}\ \overline \psi \epsilon \cr}\eqno(7)$$
and
$$\eqalign{\overline \delta \psi &= 0\cr
\noalign{\vskip 4pt}%
\overline \delta \ \overline \psi &= {i \over \sqrt{2}}\
\overline \epsilon \cr
\noalign{\vskip 4pt}%
\overline \delta p &=
 - {i g \over \sqrt{2}}\ \overline \epsilon \psi \cr}\eqno(8)$$
where $\epsilon$ and $\overline \epsilon$ are constant Grassmann parameters
of the two transformations.  In fact, note that we can use the nilpotent
nature of the fermionic variables to write the Hamiltonian in Eq. (6) also
as
$$H = {1 \over 2}\ \left( p + g \overline \psi \psi \right)^2 \eqno(9)$$
and from the transformations in Eqs. (7) and (8), it is obvious that
$$\delta \left( p + g \overline \psi \psi \right) = 0 = \overline \delta
\left( p + g \overline \psi \psi \right) \eqno(10)$$

\medskip

It is clear from Eqs. (7) and (8) that
$$\delta^2 \phi = 0 = \overline \delta^2 \phi \eqno(11)$$
for any variable, $\phi$, reflecting the fact that the generators
 of these
transformations are nilpotent as supersymmetry generators ought to be.
However, it is more interesting to note that
$$\left[ \delta_1 , \overline \delta_2 \right] \phi = 0 \eqno(12)$$
for any variable, leading to the fact that the generators of the two
supersymmetry transformations anticommute with each other, namely,
$$\left[ Q, \overline Q\  \right]_+ = 0 \eqno(13)$$
This is more like the graded algebra satisfied by the BRST and anti-BRST
generators [3] than the conventional graded algebra of Eq. (5).  Furthermore,
we note from Eqs. (7) and (8) that
$$\eqalign{<\delta \psi>\ &= - {i \over \sqrt{2}}\ \epsilon
 \not= 0\cr
\noalign{\vskip 4pt}%
<\overline \delta \ \overline \psi>\ &=  {i \over \sqrt{2}}\
\overline \epsilon
 \not= 0\cr}\eqno(14)$$
It would follow, therefore, that both the \lq\lq supersymmetries" are
spontaneously broken [4] unlike the BRST symmetries.  To understand better the
nature of these symmetries, we note that the Lagrangian for the system in
Eq. (6) is given by
$$L = {1 \over 2}\ \left( \dot q \right)^2 + i \overline \psi \dot \psi
- g \dot q \overline \psi \psi \eqno(15)$$
It is clear that since the interaction is a derivative interaction, it does
not have a natural superspace description.  In fact, note that the
invariances of the Lagrangian are given by
$$\eqalign{\delta \psi &= - {i \over \sqrt{2}}\ \epsilon\cr
\noalign{\vskip 4pt}%
\delta \overline \psi &= 0\cr
\noalign{\vskip 4pt}%
\delta \dot q &=  {i g \over \sqrt{2}}\ \overline \psi \epsilon\cr}
\eqno(16)$$
and
$$\eqalign{\overline \delta \psi &= 0\cr
\noalign{\vskip 4pt}%
\overline \delta \psi &= {i \over \sqrt{2}}\ \overline \epsilon\cr
\noalign{\vskip 4pt}%
\overline \delta \dot q &= - {i g \over \sqrt{2}}\
\overline \epsilon \psi\cr}\eqno(17)$$
These are, in fact, nonlocal transformations.

\medskip

The simple Hamiltonian of Eq. (6) can also be generalized to include
interaction with an arbitrary monomial of $p$.  In fact, it is quite
straightforward to show that the Hamiltonian
$$H = {1 \over 2}\ p^2 + f(p) \overline \psi \psi \eqno(18)$$
where $f(p)$ is an arbitrary monomial of $p$ is invariant under the
supersymmetry transformations
$$\eqalign{\delta \psi &= - {i \over \sqrt{2}}\ \epsilon\cr
\noalign{\vskip 4pt}%
\delta \overline \psi &= 0\cr
\noalign{\vskip 4pt}%
\delta p &=  {i \over \sqrt{2}}\ {f(p) \over p}\
\overline \psi \epsilon\cr}\eqno(19)$$
and
$$\eqalign{\overline \delta \psi &= 0\cr
\noalign{\vskip 4pt}%
\overline \delta \ \overline \psi &=  {i \over \sqrt{2}}\
\overline \epsilon\cr
\noalign{\vskip 4pt}%
\overline \delta p &= - {i \over \sqrt{2}}\
{f(p) \over p}\ \overline \epsilon \psi\cr}\eqno(20)$$
The Lagrangian, in this case, cannot be expressed in a simple form as in Eq.
(15), but all the discussion on the properties of the supersymmetry
transformations still hold.

\medskip

The Hamiltonian so far has been independent of the coordinates.  We can
introduce coordinates as well.  Consider, for example, the simple
Hamiltonian
$$H = {1 \over 2}\ p^2 + {1 \over 2}\ \omega^2 q^2 + m \overline \psi \psi
+ g p \overline \psi \psi \eqno(21)$$
For $g=0$ and $m = \omega$, this Hamiltonian describes the supersymmetric
oscillator [4] satisfying an algebra of the form in Eq. (5).  It is easy to
check that the Hamiltonian in Eq. (21) is invariant under the supersymmetry
transformations
$$\eqalign{\delta q &= {im \over \sqrt{2}\ \omega^2 q}\ \overline
\psi \epsilon\cr
\noalign{\vskip 4pt}%
\delta \psi &= -{i \over \sqrt{2}}\ \epsilon\cr
\noalign{\vskip 4pt}%
\delta \overline \psi  &= 0\cr
\noalign{\vskip 4pt}%
\delta p &= {i \over \sqrt{2}}\  g \overline
\psi \epsilon\cr}\eqno(22)$$
and
$$\eqalign{\overline \delta q &= {im \over \sqrt{2}\ \omega^2 q}\ \overline
\epsilon \psi\cr
\noalign{\vskip 4pt}%
\overline \delta \psi &= 0\cr
\noalign{\vskip 4pt}%
\overline \delta \ \overline \psi  &= {i \over \sqrt{2}}\
\overline \epsilon\cr
\noalign{\vskip 4pt}%
\overline \delta p &= -{i \over \sqrt{2}}\  g \overline
\epsilon \psi\cr}\eqno(23)$$
These are generalizations of the transformations in Eqs. (19) and (20) and
the generators
can be easily checked to satisfy a BRST-like algebra.  The fermionic
symmetries would appear to be spontaneously broken as discussed earlier.
 The model in Eq. (21) is, in fact, quite
interesting in that  it can be exactly diagonalized and that the true
ground state of the system is a fermionic, coherent state in contrast to
the bosonic perturbative ground state with no quantum [5].

\medskip

This construction can be extended to more complicated Hamiltonians of the
form
$$H = {1 \over 2}\ p^2 + {1 \over 2}\ \left( f(q) \right)^2 +
m \overline \psi \psi + g (p) \overline \psi \psi \eqno(24)$$
where $f(q)$ and $g(p)$ are arbitrary monomials of the coordinate and
momentum respectively.  This Hamiltonian is invariant under supersymmetry
transformations of the form
$$\eqalign{\delta q &= {im \over \sqrt{2}\ f(q) f^\prime (q)}\ \overline
\psi \epsilon \cr
\noalign{\vskip 4pt}%
\delta \psi  &= -{i \over \sqrt{2}}\ \epsilon \cr
\noalign{\vskip 4pt}%
\delta \overline \psi &= 0\cr
\noalign{\vskip 4pt}%
\delta p &= {i \over \sqrt{2}}\ {g(p) \over p}\ \overline
\psi \epsilon \cr}\eqno(25)$$
and
$$\eqalign{\overline \delta q &=
- {im \over \sqrt{2}\ f(q) f^\prime (q)}\ \overline
\epsilon \psi \cr
\noalign{\vskip 4pt}%
\overline \delta \psi &= 0\cr
\noalign{\vskip 4pt}%
\overline \delta\  \overline \psi  &= {i \over \sqrt{2}}\
\overline \epsilon \cr
\noalign{\vskip 4pt}%
\overline \delta p &= -{i \over \sqrt{2}}\ {g(p) \over p}\ \overline
\epsilon \psi  \cr}\eqno(26)$$
which satisfy a BRST-like algebra.  It  is interesting, however, to note
that the Hamiltonian in Eq. (24) is also invariant under the fermionic
transformations
$$\eqalign{\delta q &=
{1 \over \sqrt{2}}\ {1 \over f^\prime (q)}\
(m+g(p)) \overline
\psi \epsilon  \cr
\noalign{\vskip 4pt}%
\delta \psi &= -{i \over \sqrt{2}}\ (p-if(q))\epsilon\cr
\noalign{\vskip 4pt}%
\delta \overline \psi  &= 0\cr
\noalign{\vskip 4pt}%
\delta p &= -{i \over \sqrt{2}}\
(m+g(p)) \overline
  \psi \epsilon  \cr}\eqno(27)$$
and
$$\eqalign{\overline \delta q &=
{1 \over \sqrt{2}}\ {1 \over f^\prime (q)}\
(m+g(p)) \overline
\psi \epsilon  \cr
\noalign{\vskip 4pt}%
\overline \delta \psi &= 0\cr
\noalign{\vskip 4pt}%
\overline \delta\  \overline \psi  &= {i \over \sqrt{2}}\
(p+if(q)) \overline \epsilon \cr
\noalign{\vskip 4pt}%
\overline \delta p &= -{i \over \sqrt{2}}\
(m+g(p)) \overline
\epsilon \psi  \cr}\eqno(28)$$
The algebra in this case, however, is neither BRST-like nor of the kind in
Eq. (5).

\medskip

I would like to thank Prof. S. Okubo for his comments.
This work was supported in part by the U.S. Department of Energy Grant No.
DE-FG-02-91ER40685.

\vfill\eject

\noindent {\bf References}

\medskip

\item{1.} E. Witten, Nucl. Phys. {\bf B202} (1982) 253.

\item{2.} For a review of supersymmetry see P. Fayet and S. Ferrara, Phys.
Rep. {\bf 32} (1977) 249; P. Van Nieuwenhuizen, Phys. Rep. {\bf 68},
(1981) 264.

\item{3.} T. Kugo and I. Ojima, Prog. Theor. Phys. {\bf 60} (1978) 1869.

\item{4.} See for example, A. Das, Physica {\bf A158} (1989) 1.

\item{5.} S. P. Misra, \lq\lq Phase Transitions in Quantum Field Theory",
hep-ph/9212287.

\end